\documentclass[11pt]{article}
\usepackage{amsmath}
\usepackage{amsfonts}
\usepackage{graphicx}
\usepackage{enumerate}
\usepackage[sort,compress]{cite}
\usepackage[papersize={8.5in,11in},margin=1in]{geometry}

\begin{document}
\begin{center}
  \Large\bf{
Sampling, feasibility, and priors in Bayesian estimation
}
\end{center}

\begin{center}
Alexandre J. Chorin$^{1,2}$, Fei Lu$^{1,2}$, Robert N. Miller$^3$,  Matthias Morzfeld$^{1,2}$, Xuemin Tu$^4$
\vspace{2mm}

$^1$Department of Mathematics, University of California, Berkeley, CA \\
$^2$Lawrence Berkeley National Laboratory, Berkeley, CA \\
$^3$College of Oceanic and Atmospheric Sciences, Oregon State University, Corvallis, OR \\
$^4$Department of Mathematics, University of Kansas, Lawrence, KS \\
\end{center}
\vspace{1mm}

\begin{center}
DEDICATED TO PETER LAX, ON THE OCCASION OF HIS 90th BIRTHDAY, WITH FRIENDSHIP 
AND ADMIRATION
\end{center}

\vspace{3mm}

\begin{center}
\emph{Abstract}
\end{center}

Importance sampling algorithms are discussed in detail, with an emphasis
on implicit sampling, and applied to data assimilation via particle filters. 
Implicit sampling makes it possible to use the data to find high-probability samples at relatively low cost, making the assimilation more
efficient.  
A new analysis of the feasibility of data assimilation is presented,
showing in detail why feasibility depends on the Frobenius norm
 of the covariance matrix of the noise and not on the number of variables. A 
 discussion of the convergence of particular particle filters follows. 
A major open problem in numerical data assimilation is the determination
of appropriate priors; a progress report on recent work on this
problem is given. 
The analysis highlights the need for a    
careful attention both to the data and to the physics in data assimilation problems.

\section{Introduction}
Bayesian methods for estimating parameters and states in complex systems are widely used in science and engineering; they combine a prior distribution of the quantities of interest, often generated by computation, with data from observations, to produce a posterior distribution from which reliable inferences can be made. Some recent applications of these methods, for example in geophysics, involve more unknowns, larger data sets, and more
nonlinear systems than earlier applications, and present new challenges in their use
(see, e.g.\cite{Bocquet2010,PeterJan2009,Fournier2010,Stuart2010,Doucet2001} for recent reviews of Bayesian methods in engineering and physics).

The emphasis in the present paper is on data assimilation via particle filters, which 
requires effective sampling. 
We give a preliminary discussion of importance sampling
and explain that implicit sampling\cite{Morzfeld2011,atkins,chorintupnas,chorin2010}
is an effective importance sampling method.
We then present  implicit sampling methods
for calculating a posterior distribution of the unknowns of interest,
given a prior distribution and a distribution of the observation errors,
first in a parameter estimation problem, then in a
data assimilation problem where the prior is generated by solving
stochastic differential equations with a given noise.
Conditions for data assimilation to be feasible in principle
and their physical interpretation are discussed (see also\cite{CM13}).
A linear convergence analysis for data assimilation methods
shows that Monte Carlo methods converge
for many physically meaningful data assimilation problems,
provided that the numerical analysis is appropriate and that the size of
the noise is small enough in the appropriate norm, even when the number of variables is very large.
The keys to success are a careful use of data and a careful attention
to correlations.

A very important open problem in the practical application of data assimilation
methods is the difficulty in finding suitable error models, or priors.
We discuss a family of problems
where priors can be found, with an example. Here too
a proper use of data and a careful attention to correlations are
the keys to success.

\section{Importance sampling}
\label{sec:importance sampling}

Consider the problem of sampling a given probability density function (pdf)  $f$ on the computer, i.e., given a probability density
function (pdf) for an $m$-dimensional random vector,
construct a sequence
of vectors $X_1,X_2, \dots$, which can pass statistical independence tests, and whose histogram converges to the graph of $f$, in symbols, find $X_i \sim f.$
We discuss this problem in the context of numerical quadrature.
Suppose one wishes to evaluate the integral $$I=\int g(y)f(y)dy,$$ where
$g(y)$ is a given function and $f(y)$ is a pdf, i.e., $f(y) \ge 0$ and $\int f\,dy=1.$
The integral $I$ equals $E[g(x)]$,
where $x$ is a random variable with
pdf $f$ and $E[\cdot]$ denotes an expected value.
This integral can be approximated via the law of large numbers as
\begin{equation*}
I = E\left[g(x)\right] \approx \frac {1}{M} \sum_{i=1}^M g(X_i),
\end{equation*}
where $X_i \sim f$
and $M$ is a large integer (the number of samples we draw).
The error is of the order of $\sigma(g(x))/M^{1/2}$,
where $\sigma(\cdot)$ denotes the standard deviation of the variable in the parentheses.
This assumes that one knows how to find the samples $X_i$, which in
general is difficult. One way to proceed is
importance sampling (see, e.g.,\cite{KalosWhitlock,ChorinHald}):
introduce an auxiliary pdf $f_0$, often called an importance function,
which never vanishes unless $f$ does, and
which one knows how to sample, and rewrite the integral as
\begin{equation*}
	I=\int g(y)\frac{f(y)}{f_0(y)}\;f_0(y)dy = E\left[g(x^*)w(x^*)\right]
\end{equation*}
where the pdf of $x^*$ is $f_0$ and $w(x^*)=f(x^*)/f_0(x^*)$.
The expected value can then be approximated by
\begin{equation}
\label{quad0}
I \approx \frac {1}{M} \sum_{i=1}^M g(X^*_i)w(X^*_i) 
\end{equation}
where $X^*_i \sim f_0$ and the $w(X_i^*)=f(X_i^*)/f_0(X_i^*)$ are the sampling weights. 
The error in this approximation is of order $\sigma(gw)/M^{1/2}$
(here the standard deviation is computed with respect to the pdf $f_0$). 
The sampling weights
make up for the fact that one is sampling the wrong pdf.

\begin{figure}[tb]
\centering\includegraphics[width=.5\textwidth]{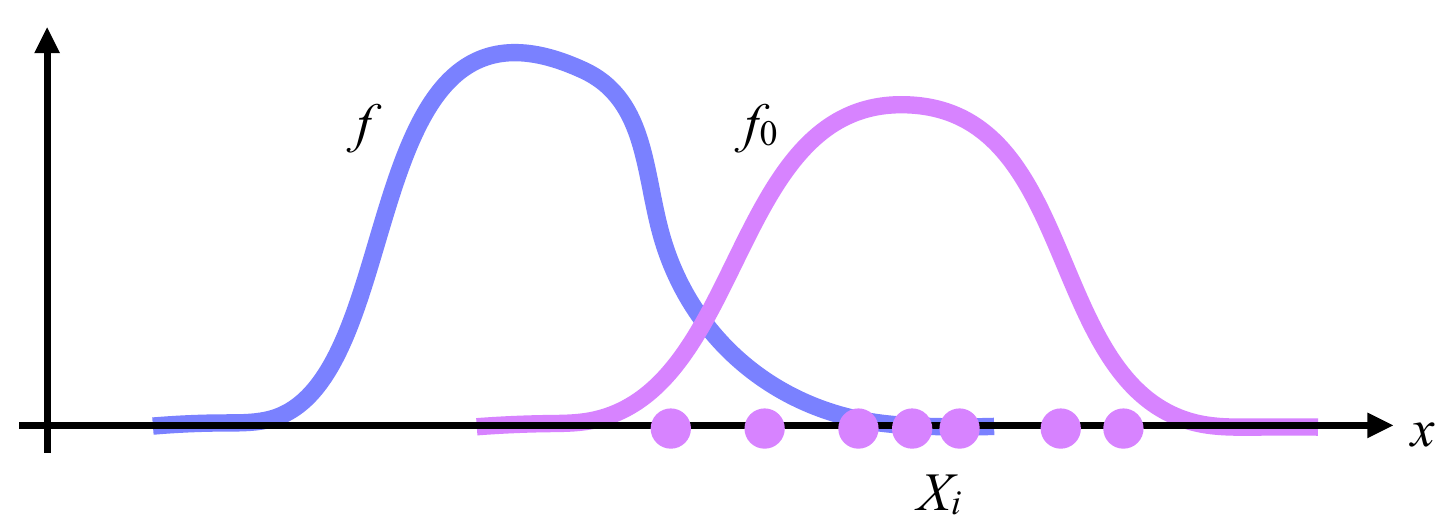}
\caption{Poor choice of importance function: the importance function (purple)
is not close to the pdf we wish to sample (blue).}
\label{fig:ImportanceFunction}
\end{figure}

Importance sampling can work well if $f$ and $f_0$ are fairly close to each other.
Suppose they are not (see figure~\ref{fig:ImportanceFunction}).
Then many of the samples $X_i$ (drawn from the importance function)
fall where $f$ is negligible and their sampling weight is small.
The computing effort used to generate such samples is wasted
since the low-weight samples contribute little to the approximation of the expected value.
In fact, one can define an effective number of samples \cite{Doucet2001,GordonReview,Weare2013,liuchen1995} as
\begin{equation*}
	M_\text{eff} = \frac{M}{R},\quad R=\frac{E\left[w^2\right]}{E\left[w\right]^2},
\end{equation*}
so that in particular, if all the particles have weights $1/M$, $M_\text{eff}=M$. 
The effective number of samples approximates the equivalent number of
independent, unweighted samples one has after importance sampling
with $M$ samples.
Importance sampling is efficient if $R$ is smaller than $M$.
For example, suppose that 
you find that one weight is Rmuch larger than all the others.
Then $R\approx M$, so that one has effectively one sample
and this sample does not necessarily have a high probability.
In this case, the importance sampling scheme has ``collapsed''.
To find $f_0$ that prevents a collapse, 
one has to know the shape of $f$ quite well,
in particular know  the region where $f$ is large.
In some of the examples below,
the whole problem is to estimate where a particular pdf is large,
and it is not realistic to assume that this is known in advance.

\begin{figure}[tb]
\centering\includegraphics[width=.4\textwidth]{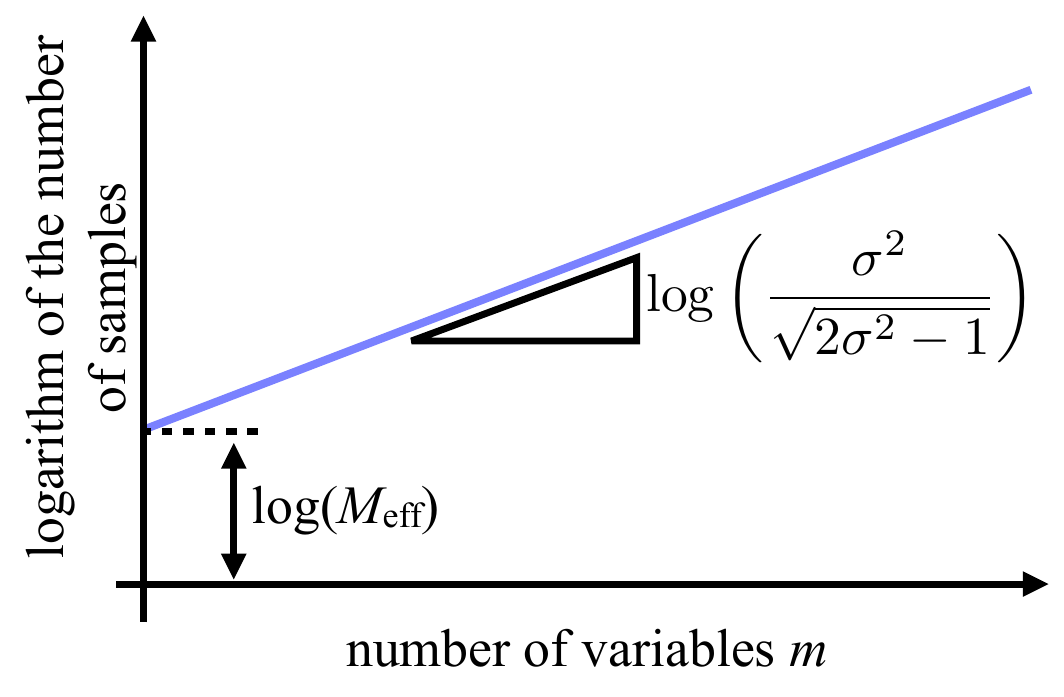}
\caption{The number of samples required increases exponentially with the number of variables.}
\label{fig:ImportanceFunctionDimension}
\end{figure}

As the number of variables increases,
the value of a carefully designed importance function also increases.
This can be illustrated by an example.
Suppose that $f = \mathcal{N}(0,I_m)$, where $I_m$ is the identity matrix
of order $m$ (here and below we denote a Gaussian with
mean $\mu$ and covariance matrix $Q$ by $\mathcal{N}(\mu,Q)$).
To sample this Gaussian we pick a Gaussian importance function $f_0 = \mathcal{N}(0,\sigma^2I_m)$, $\sigma >1/2$. The importance function has a shape similar to that of the function
we wish to sample and $f_0$ is large where $f$ is large (for moderate $\sigma$).
Nonetheless, sampling becomes increasingly costly as the number of components of $x$ increases.
We find that 
\begin{equation*}
	w(x) =\sigma^{m} \exp\left(-\frac{1}{2} \frac{\sigma^2-1}{\sigma^2}\; x^Tx\right),
\end{equation*}
where the superscript $T$ denotes a transpose, so that
\begin{equation*}
	E\left[w(x)\right] = 1,\quad E\left[w(x)^2\right] = \left(2\sigma^2-1\right)^{-\frac{m}{2}},
\end{equation*}
which gives
\begin{equation*}
	R = \left(\frac{\sigma^2}{ \sqrt{2\sigma^2-1} }\right)^{m}.
\end{equation*}
The number of samples required for a given $M_\text{eff}$
thus grows exponentially with the number of variables $m$:
\begin{equation*}
	\log(M) = m\; \log\left(\frac{\sigma^2}{ \sqrt{2\sigma^2-1} }\right) + \log(M_\text{eff}).
\end{equation*}
The situation is illustrated in figure~\ref{fig:ImportanceFunctionDimension}.
As the number of variables increases, the cost of sampling, measured by the number of samples
required to obtain the pre-specified number of effective samples,
grows exponentially with the number of variables for any $\sigma > 1/2$ except $\sigma=1$; see also
the discussion in \cite{mcbook}. 

\section{Implicit sampling}
\label{secc}
Implicit sampling is a general numerical method for constructing effective importance
functions\cite{Morzfeld2011,atkins,chorintupnas,chorin2010}.
We write the pdf to sample as $f= e^{-F(x)}$ and, temporarily and for simplicity,
assume that $F(x)$ is convex.
The idea is to write
$x$ as a one-to-one and onto function of an easy-to-sample reference variable $\xi$
with pdf $g \propto e^{-G(\xi)}$, were $G$ is convex.
To find this function, we first find the minima
of $F,\,G$, call them $\phi_F = \min F, \phi_G = \min G$,
and define $\phi=\phi_F-\phi_G$.
Then we define a mapping $\psi: \xi \rightarrow x$
such that $\xi$ and $x$ satisfy the equation
\begin{equation}
F(x)-\phi=G(\xi).
\label{map}
\end{equation}
With our convexity assumptions a one-to-one and onto map exists, in fact there are many, because equation (\ref{map}) is underdetermined (it is a single equation that connects the $m$ elements of $X_i$ to the $m$ elements of $\Xi_i$, where $m$ is the number of variables).    
To find samples $X_1,X_2,\dots$, first find samples
$\Xi_1,\Xi_2,\dots$ of $\xi$, and for each $i=1,2,\dots$, solve~(\ref{map}).
A sample $X_i$ obtained this way has a high probability (with a high probability) for the following reasons.
A sample $\Xi_i$ of the reference density $g$ is likely to lie near the minimum of $G$,
so that the difference $(G(\Xi_i)- \phi_G)$ is likely to be small.
By equation (\ref{map}) the difference $(F(X_i)-\phi_F)$ is also likely to be small and
the sample $X_i$ is in the region where $f$ is large.
Thus, by solving~(\ref{map}),
we map the high-probability region of the reference variable $\xi$ to the high-probability region of $x$, so that
one obtains a high-probability sample $X_i$
from a high-probability sample $\Xi_i$.

The mapping $\psi$ defines the importance function implicitly by the solutions of~(\ref{map}):
\begin{equation*}
	f_0(x(\xi)) = g(\xi) \left| \det\left(\frac{\partial\xi}{\partial x} \right)\right|.
\end{equation*}
A short calculation shows that the sampling weight $w(X_i)$ is
\begin{equation}
w(X_i) \propto e^{-\phi}\left| J(X_i) \right|,
\label{totw}
\end{equation}
where $J$ is the Jacobian $\det(\partial x /\partial \xi)$.
The factor $e^{-\phi}$, common to all the weights, is immaterial here,
but plays an important role further below.

We now give an example of a map $\psi$ that makes this construction easy to implement.
We assume here and for the rest of the paper that the reference variable $\xi$
is the Gaussian $\mathcal{N}(0,I_m)$, where $m$ is the dimension of $x$.
With a Gaussian $\xi$, $\phi_G=0$ and equation (\ref{map}) becomes
\begin{equation}
\label{eq:ImplicitEquation}
	F(x)-\phi=\frac{1}{2}\xi^T\xi.
\end{equation}
We can find a map that satisfies this equation by looking
for solutions in a random direction $\eta = \xi/(\xi^T\xi)$, i.e. use a mapping $\psi$ such that
\begin{equation}
\label{matti}
x=\mu+\lambda L \eta,
\end{equation}
where $\mu=\mbox{argmin}\,F$ is the minimizer of $F$, $L$ is a given matrix, and $\lambda$ is a scalar that depends on $\xi$. Substitution of the above mapping into (\ref{eq:ImplicitEquation}) gives a scalar equation in the single variable $\lambda$ (regardless of the dimension of $x$). The matrix $L$ can be chosen to optimize the distribution of the sampling weights. For example, if $L$ is a square root of the inverse of the Hessian of $F$ at the minimum, then 
the algorithm is affine invariant, which is important in applications with multiple scales\cite{G10}.
Equation (\ref{matti}) can be readily solved and the resulting Jacobian is easy to calculate (see \cite{Morzfeld2011} for details).

Other constructions of suitable maps $\psi$ are presented in e.g.\cite{chorin2010,GLM15}
and some of these are analyzed in\cite{GLM15}.
With these constructions,
often the most expensive part of the calculation is finding the minimum of $F$.
Note that this needs to be done only once for each sampling problem,
and that finding the maximum of $f$ is an unavoidable part of any
useful choice of importance function, explicitly or implicitly.
Addressing the need for the optimization explicitly
has the advantage that effective optimization methods can be brought to bear.
Connections between optimization and (implicit) sampling also have been pointed out in\cite{atkins,MTWC2015}.
In fact, an alternate derivation of implicit sampling can be given
by starting from maximum likelihood estimation, followed by a randomization that
removes bias and provides error estimates.

We now relax the assumption that $F$ is convex. If $F$ is $U$-shaped, the above construction works without modification. A scalar function $F$ is $U$-shaped if it is piecewise differentiable, its first derivative vanishes at a single point which is a minimum, $F$ is strictly decreasing on one side of the minimum and strictly increasing on the other, and $F(x)\rightarrow \infty$ as $\left|x\right| \rightarrow \infty$; in the general case, $F$ is $U$-shaped if it has a single minimum and each intersection of the graph of the function $y = F(x)$ with a vertical plane through the minimum is $U$-shaped in the scalar sense. If $F$ is not $U$-shaped, but has only one minimum, one can replace it by a $U$-shaped approximation, say $F_0$, and then apply implicit sampling as above. The bias created by this approximation can be annulled by reweighting \cite{chorin2010}. If there are multiple minima, one can represent
$f$ as a mixture of components whose logarithms are $U$-shaped,
and then pick as a reference pdf $g$ a cross-product of a Gaussian and
a discrete random variable. However, the decomposition into a suitable
mixture can be laborious (see \cite{brad}).

\section{Beyond implicit sampling}
\label{secbeyond}

Implicit sampling produces a sequence of samples that lie in the high-probability domain and are weighted by a Jacobian. It is natural to wonder whether
one can absorb the Jacobian into the map $\psi : \xi \rightarrow x$ and obtain ``optimal" samples that need no weights
(here and below, optimal refers to minimum variance of the weights, i.e.~an optimal sampler is one
with a constant weight function).
If a random variable $x$ with pdf $f$ is a one-to-one function of a variable $\xi$ with pdf $g$, then
\begin{equation}
f(x(\xi))=g(\xi)J(\xi),
\label{marzouk}
\end{equation}
where $J$ is the Jacobian of the map $\psi$.
One can obtain samples of $x$ directly (i.e.~without weights)
by solving~(\ref{marzouk}).
Notable efforts in that direction can be found in \cite{Tabak,Marzouk2012}. 

However, optimal samplers can be expensive to use; 
the difficulties can be demonstrated by the following construction.
Consider a one-variable pdf $f$ with a convex $F=-\log\,f$.
Find the maximizer $z$ of $f$, with maximum $M_f=f(z)$, and let $g$ be the pdf of a
reference variable $\xi$, with its maximizer also at $z$ and maximum $M_g=g(z)$.
To simplify the notations, change variables so that $z=0$.
Then construct a mapping $\psi : \xi \rightarrow x$ as follows.
Solve the differential equation
\begin{equation}
\label{ode}
\frac{du}{dt}=\frac{g(t)}{f(u)},
\end{equation}
where $t$ is the independent variable and $u$ is a function of $t$,
in the $t$-interval $[0,\xi]$, with initial condition $u=0$ when $t=0.$
Define the  map from $\xi$ to $x$ by $x=u(\xi)$; then $f(x)=g(\xi)J$,
where $J=|du/dx|$ evaluated at $t=\xi$. It is easy to see that under the assumptions made, this map is one-to-one and onto. We have achieved optimal sampling.

The catch is that the initial value of $g(t)/f(u)$ is $M_g/M_f$, which requires
that the normalization constants of $f$ and $g$ be known.
In general
the solution of the differential equation does not depend linearly on the
initial value of $g(0)/f(u(0))$, so that unless $M_f$ and $M_g$ are known,
the samples are in the wrong place while weights to remove the resulting bias are unavailable.
Analogous conclusions were reached in \cite{Doucet} for a different sampling scheme.
In the special case where $F=-\log\,f$ and $G=-\log\,g$ are both quadratic functions
the constants $M_f,M_g$ can be easily calculated, but under these conditions
one can find a linear map that satisfies equation (\ref{marzouk}) and implicit sampling becomes identical to optimal sampling.
The elimination of all variability in the weights in the nonlinear case 
is rarely worth the cost of computing the normalization constants.

Note that the normalization constants are not needed for implicit sampling.
The mapping~(\ref{matti}) for solving~(\ref{map}) is well-defined even when the
normalization constants of $f$ and $g$ are unknown.
The reason is that if one multiplies the functions $f$ or $g$
by a constant, the logarithm of this constant is added both to
$F$ ($G$) and to $\phi_F$ ($\phi_G$) and cancels out.
Implicit sampling simplifies equation~(\ref{marzouk})
by assuming that the Jacobian is a constant.
It produces weighted samples where the weights are known only up to a constant,
and this indefiniteness can be removed by normalizing the weights so that their sum equals
1.

\section{Bayesian parameter estimation}
\label{sec:parameters}
We now apply implicit sampling to Bayesian estimation. For the sake of clarity we first explain how to use Bayesian estimation to determine physical parameters needed in the
numerical modeling of physical processes, given noisy observations of these processes. This discussion explains how a prior and data combine to create a posterior proability density that is used for inference. In the next section we extend the methodology to data assimilation, which is our main focus.

Suppose one wishes to model the diffusion of some material in a given medium.
The density of the diffusing material
can be described by a diffusion equation, provided the diffusion coefficients
$\theta \in \mathbb{R}^{m}$
can be determined. Given the coefficients $\theta$,
one can solve the equation and determine the values of the density at a
set of points in space and time. The values of the density at these
points can be measured experimentally, by some method with an error $\eta$
whose probability density is assumed known; once the measurements $d \in
\mathbb{R}^{k}$ are made,
this assigns a probability to $\theta$.
The relation between $d$ and $\theta$ can be written as
\begin{equation}
\label{eq:data}
d=h(\theta )+\eta ,
\end{equation}
where $h:\mathbb{R}^{k}\rightarrow \mathbb{R}^{m}$ is a generally nonlinear function, and $\eta\sim p_\eta(\cdot)$ is a random variable with  known pdf that represents the uncertainty in the measurements.
The evaluation of $h$ often requires a solution of a differential equation and can be expensive.
In the Bayesian approach, one assumes that information about the parameters is available in form of a prior $p_0(\theta)$. This prior and the likelihood $p(d\vert \theta)=p_\eta(d-h(\theta) )$, defined by (\ref{eq:data}), are combined via Bayes' rule to give the posterior pdf, which is the probability density function for the parameters $\theta$ given the data $d$,
\begin{equation}
p(\theta |d)=\frac{1}{\gamma (d)}p_{0}(\theta )p(d|\theta ),  \label{eq:post}
\end{equation}
where  $\gamma (d)=\int p_{0}(\theta )p(d|\theta )d\theta $ is a normalization constant
(which is hard to compute).

In a Monte Carlo approach to the determination of the parameters,
one finds samples of the posterior pdf.
These samples can, for example, be used to approximate the expected value
\begin{equation*}
	E[\theta\vert d]=\int \theta\, p(\theta \vert d)\,d\theta,
\end{equation*}
which is the best approximation of $\theta$ in a least squares sense (see, e.g.\cite{ChorinHald}),
and a error estimate can also be computed, see also\cite{Stuart2010}.

When sampling the posterior  $p(\theta |d)$,
it is natural to set the importance function
equal to the prior probability $p_0(\theta)$ and then weight the samples
by the likelihood $p(d|\theta)$.
However, if the data are informative, i.e., if the measurements
$d$ modify the prior significantly, then
the prior and the posterior can be very different and this importance sampling scheme is ineffective.
When implicit sampling is used to sample the posterior,
then the data are taken into account already when generating the samples,
and not only in the weights of the samples.

As the number of variables increases, the
prior $p_0$ and the posterior $p(\theta| d)$ may become nearly mutually singular.
In fact, in most practical problems these pdfs are negligible outside a small spherical domain
so that the odds that the prior and the posterior have a significant overlap decrease
quickly with the the number of variables, making
implicit sampling increasingly useful
(see\cite{MTWC2015} for an application of implicit sampling to parameter estimation).

\section{Data assimilation}
We now apply Bayesian estimation via implicit sampling to data assimilation,
in which one makes inferences from an unreliable time-dependent approximate model that defines a prior,
supplemented by a stream of noisy and/or partial data. We assume here that the model is
a discrete recursion of the form
\begin{equation}
\label{sde}
x^{n+1}=f(x^n)+w^n,
\end{equation}
where $n$ is a discrete time, $n=0,1,2,\dots $, the set of $m$-dimensional vectors $x^{0:n} = (x^0,x^1,\dots,x^n,\dots)$ are the state
vectors we wish to estimate,
$f(\cdot)$ is a given function, and $w^n$ is a Gaussian random vector $\mathcal{N}(0,Q).$
The model often represents a discretization of a
stochastic differential equation.
We assume that $k$-component data vectors $b_n,\, n=1,2, \dots$ are
related to the states $x^n$ by
\begin{equation}
b^n=h(x^n)+\eta^n,
\label{obs}
\end{equation}
where $h(\cdot)$ is the (generally nonlinear) observation function and
$\eta^n$ is a $\mathcal{N}(0,R)$ Gaussian vector
(the $\eta^n$ and $w^n$ are independent of $\eta^k$ and $w^k$ for $k < n$ 
and also independent of each other).
In addition, initial data $x^0$, which may be random, are given
at time $n=0$.
For simplicity, we consider in this section the problem of estimating
the state $x$ of the system rather than estimating parameters in the
equations as in the previous section.
Thus, we wish to sample the conditional pdf
$p(x^{0:n}|b^{1:n})$, which describes the probability of the sequence of states between $0$ and $n$
given the data $b^{1:n}=(b^1,\dots,b^n)$.
The samples of this conditional pdf
are sequences $X^0,X^1,X^2,\dots$, usually referred to as ``particles".
The conditional pdf satisfies the recursion
\begin{equation}
p(x^{0:n+1}|b^{1:n+1})=p(x^{0:n}|b^{1:n})\frac{p(x^{n+1}|x^n)p(b^{n+1}|x^{n+1})}{p(b^{n+1}|b^{1:n})},
\label{recursion}
\end{equation}
where $p(x^{n+1}|x^n)$ is determined by equation (\ref{sde}) and
$p(b^{n+1}|x^{n+1})$ is determined by equation (\ref{obs}).
(see e.g. \cite{Doucet2001}). We wish to sample the conditional pdf recursively,
time step after time step,
which is natural in problems where the data are obtained sequentially.
To do this we use an importance function $\pi$ of the
form
\begin{equation}
\pi(x^{0:n+1}|b^{1:n+1})=\pi_0(x^0) \prod_{k=1}^{n+1}\pi_k(x^k|x^{0:k-1},b^{1:k}).
\label{factor}
\end{equation}
where the $\pi_k$ are increments of the importance function.
The recursion (\ref{recursion}) and the factorization (\ref{factor}) yield
a recursion for the weight $W_j^{n+1}$ of the $j$-th particle,
\begin{equation}
W_j^{n+1}= \frac{p(x^{0:n+1}|b^{1:n+1})}{\pi(x^{0:n+1}|b^{1:n+1})} = W_j^n \frac {p(X_j^{n+1}|X_j^n)p(b^{n+1}|X^{n+1}_j)}{\pi_{n+1}(X_j^{n+1}|X^{0,n},b^{1:n})}
\label{werecur}.
\end{equation}
With this recursion, the task at hand is to pick a suitable update $\pi_k(x^k|x^{0:k-1},b^{1:k})$
for each particle, to sample $p(X_j^{n+1}|X_j^n)p(b^{n+1}|X^{n+1}_j)$,
and to update the weights so that the pdf
$p(x^{0:n+1}|b^{1:n+1})$ is well-represented.
Thus, the solution of the discrete model~(\ref{sde})
plays the same role in data assimilation as the prior in the parameter estimation problem of section~5.

In the SIR (sequential importance sampling) filter one picks
\begin{equation*}
	\pi_{n+1}(x^{n+1}|x^{0:n},b^{1:n+1}) = p(x_j^{n+1}|X_j^{n-1}),
\end{equation*}
and the weight is $W_j^{n+1} \propto W_j^n\,p(b^{n+1}|X^{n+1}_j)$.
Thus, the SIR filter proposes samples by the model~(\ref{sde})
and determines their probability from the data~(\ref{obs}).

In implicit data assimilation one
samples $\pi_{n+1}(x^{n+1}|x^{0:n},b^{1:n+1})$ by implicit sampling, thus taking the most recent data $b^{n+1}$ into
account for generating the samples. The weight of each particle is
given by equation (\ref{werecur})
\begin{equation*}
	W_j^{n+1} = W_j^{n} \exp(-\phi_j^{n+1}) J(X^{n+1}_j),
\end{equation*}
where $\phi_j^{n+1} = \text{argmin}\; F_j^{n+1}$ is the minimum of
\begin{equation*}
	F_j^{n+1}(x^{n+1}) = -\log\left( p(x^{n+1}|X_j^n) p(b^{n+1}\vert x^{n+1})\right).
\end{equation*}
Thus, a minimization is required for each particle at each step.

The ``optimal'' importance function\cite{Doucet,OptimalImportanceFunction,liuchen1995}
is defined by~(\ref{factor}) with
\begin{equation*}
	\pi_{n+1}(x^{n+1}|x^{0:n},b^{1:n+1}) = p(x^{n+1}|x^n,b^{n+1}),
\end{equation*}
and its weights are
\begin{equation*}
	W_j^{n+1}=W_j^n \; p(b^{n+1}|X_j^n).
\end{equation*}
This choice of importance functions is optimal
in the sense that it minimizes the variance
of the weights at the $n+1$ step given the data and
the particle positions at the previous step.
In fact, this importance function is optimal over all importance functions
that factorize as in~(\ref{factor}), and in the sense
that the variance of the weights $\text{var}(w^n)$
is minimized (with expectations computed with respect
to $\pi(x^{0:n+1}|b^{0:n+1})$), see\cite{SBM15}.
In a linear problem where all the noises are Gaussian,
the implicit filter and the optimal filter coincide\cite{chorintupnas,Morzfeld2011,ME}.
When a problem is nonlinear, the optimal filter may be hard to implement,
as discussed above. 

A variety of other constructions is
available (see e.g. \cite{brad,Weare2013,Weare2012,vanLeeuwen,Ades}).
The SIR and optimal filters
are the two extreme cases, in the sense that one of them makes no use of the data
in finding samples while the other makes maximum use of the data.
The optimal filter becomes impractical for nonlinear problems
while the implicit filter can be implemented at a reasonable cost.
Implicit sampling thus balances the use of data in sampling
and the computational cost.

\section{The size of a covariance matrix and the feasibility of data assimilation}

Particle filters do not necessarily succeed in assimilating data. There are
many possible causes- for example, the recursion (\ref{sde}) may be unstable, or the data may be inconsistent. The most frequent cause of failure is filter collapse, in which only one particle has a
non-negligible weight, so that there is effectively only one particle, which is not sufficient
for valid inferences. In the next section we analyze how particle collapse
happens and how to avoid it. In preparation for this discussion, we discuss here
the Frobenius norm of a covariance matrix, its geometrical significance,
and physical interpretation.

The effectiveness of a sampling algorithm depends on the pdf that it samples.
For example, suppose that the posterior you want to sample
is in one variable, but has a large variance.
Then there is a large uncertainty in the state
even after the data are taken into account,
so that not much can be said about the state.
We wish to assess the conditions under which the posterior can
be used to draw reliable conclusions about the state,
i.e.~we make the statement ``the uncertainty is not too large'' quantitative.
We call sampling problems with a small enough uncertainty ``feasible in principle''.
There is no reason to attempt to solve a sampling problem that is not feasible in principle.

Our analysis is linear and Gaussian and we allow the number of variables to be large.
in multivariate Gaussian problems, feasibility requires that a suitable norm
of the covariance matrix be small enough. 
This analysis is inspired by geophysical applications
where the number of variables is large, but the nonlinearity may be mild; 
parts of it were 
presented in\cite{CM13}.

We describe the size of the uncertainty in multivariate problems
by the size the $m\times m$ covariance matrix $P=(p_{ij})$, which we measure
by its Frobenius norm
\begin{equation}
\vert\vert P\vert\vert_F=\left(\sum_{i=1}^m \sum_{j=1}^m p_{ij}^2\right)^{1/2}=\left(\sum_{k=1}^m \lambda_k^2\right)^{1/2},
\end{equation}
where the $\lambda_k, \,k=1,\dots,m$ are the eigenvalues of $P$.
The reason for this choice is the geometric significance of the Frobenius norm.
Let $x\sim \mathcal{N}(\mu,P)$ be an $m$-dimensional Gaussian variable,
and consider the distance between a sample of $x$ and
its most likely value $\mu$
\begin{equation*}
	r = \vert\vert x-\mu \vert\vert_2=\sqrt{(x-\mu)^T(x-\mu)}.
\end{equation*}
If all eigenvalues of $P$ are $O(1)$ and if $m$ is large,
then $E[r]$, the expected value of $r$, is $O(\vert\vert P\vert\vert_F)$ and $var(r) = O(1)$, i.e.~the samples
concentrate on a thin shell, far from their most likely value.
Different parts of the shell may correspond to physically
different scenarios (such as ``rain'' or ``no rain'').
Moreover, the expected value and variance
of the distance distance $r=\sqrt{y}$ of a sample to its most likely value
are bounded above and below by multiples of $||P||_F$ (see\cite{CM13}).
If one tries to estimate the state of a system with a large $\vert\vert P\vert\vert_F$,
the Euclidean distance between a sample and the resulting estimate is likely to be large,
making the estimate useless.
For a feasible problem we thus require that $\vert\vert P\vert\vert_F$
not be too large. How large ``large'' can be depends on the problem.

In\cite{CM13} and in earlier work\cite{Bickel,BickelBootstrap,Bickel2} on the feasibility of particle filters,
the Frobenius norm of $P$ was called the ``effective dimension''. This terminology is confusing and we abandon it here.
We show below that 
$\vert\vert P\vert\vert_F$ quantifies the strength
of the noise. We define the effective dimension of the noise by
\begin{equation}
\label{effDim2}
	\min \ell \in \mathbb{Z} : \sum_{j=1}^{\ell} \lambda_j^2 \ge (1-\epsilon) \sum_{j=1}^{\infty} \lambda_j^2,
\end{equation}
where $\epsilon$ is a predetermined small parameter.
A small $\vert\vert P\vert\vert_F$ can imply a small $\ell$,
however the reverse is not necessarily true and $\vert\vert P\vert\vert_F$ can be small,
even though $\ell$ is large
(for example if $x\sim\mathcal{N}(0,I_m/\sqrt{m})$) and vice versa.
There are small dimensional problems with a large noise (large $||P||_F$)
which are not feasible in principle, and there are large dimensional problems with a small
noise which are feasible in principle (small $||P||_F$).
Estimation based on a posterior pdf is feasible in principle only if
$\vert\vert P\vert\vert_F$ is small.

We now examine the relations between $||P||_F$, the effective dimension $\ell$, and the correlations between the components of the noise (i.e., the size
of the off-diagonal terms in $P$). We assume that our random variables
are point values $x_i$ of a stationary stochastic process on the real line, measured at the points $ih$ of the finite interval $[0,1]$,
where $i=1,2,\dots,m$ and $h=1/m$. As the mesh is refined, $m$ increases. 
The condition for the discrete problem to be feasible is that the Frobenius 
norm of its covariance matrix be bounded. Let the covariance matrix of the
continuous process be $k(x,x')=k(x-x')$; the corresponding covariance
operator is defined by $$K\varphi=\int_0^1 k(x,x')\varphi(x')dx'$$ for
every function $\varphi=\varphi(x)$. The eigenvalues of $K$ can be approximated 
by the eigenvalues of $P$ multiplied by $h$ (see \cite{Rasmussen}). 
The Frobenius norm of $K$, $$\int_0^1 \int_0^1\,k^2\,dx\,dx',$$ describes the distance of a sample to its most likely value in the $L^2$ sense,
as can be seen 
by following the same steps as in\cite{CM13}, replacing the Euclidean norm
by the appropriate grid-norm, i.e.~$||x||_2^2 = \sum_{k=1}^m h\,x_k^2$.

We consider a family of Gaussian stationary processes
with differing correlation structures. The discussion is simplified if we
keep the energy $e$ defined by 
\begin{equation*}
e = \int_{-\infty}^{\infty}k(y,y')^2 \,dy'.
\label{hu}
\end{equation*}
equal to 1 for 
all the members of the family (so that all the members of the family are
equally noisy); that $e$ is an energy follows from Khinchin's theorem (see e.g. \cite{ChorinHald}). An example of such a family is the family of zero-mean
processes with 
\begin{equation*}
k(x,x') = \pi^{-\frac{1}{4}}L^{-\frac{1}{2}}\, \exp\left(-\frac{(x-x')^2}{2L^2}\right),
\end{equation*}
where $L$ is the correlation length. The infinite limits in the definition of $e$ make the
calculations easier, and are harmless as long as $L$ is less than about $1/2$.
The elements $p_{ij}$ of the discrete matrix $P$ are 
$p_{ij} = k(ih,jh)$, $i,j = 1,\dots,m$.

\begin{figure}[tb]
\centering\includegraphics[width=.9\textwidth]{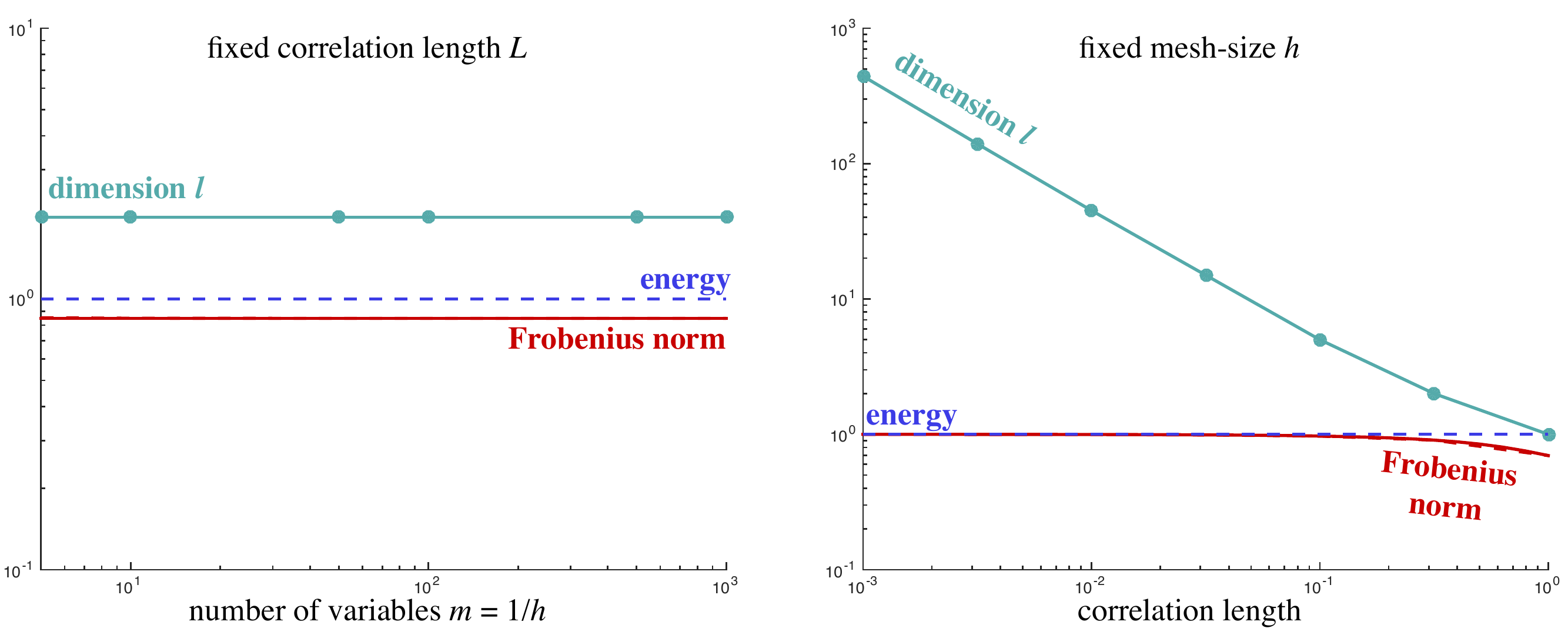}
\caption{Norms of covariance matrices and dimension $\ell$ for a family
of Gaussian processes with constant energy as a function
of the mesh size $h$ (left) and as a function of the correlation length $L$ (right).}
\label{fig:EffDimExample}
\end{figure}

In the left panel of figure~\ref{fig:EffDimExample}
we demonstrate that for a fixed correlation length $L=0.1$,
the Frobenius norm
\begin{equation*}
\vert\vert P \vert\vert_F = \left(\sum_{k=1}^m(h\,\lambda_k)^2\right)^\frac{1}{2},
\end{equation*}
where the $\lambda_k$ are the eigenvalues of the covariance matrix $P$,
is independent of the mesh (for small enough $h$) and,
therefore independent of the discretization.
In the calculation of the 
dimension $\ell$ in~(\ref{effDim2}), we use $\epsilon = 5\%$.
In the figure, we plot the Frobenius norms of  
the $m\times m$ matrices $P$ as the purple and red) dashed lines.
The solid line is the Frobenius of the covariance operator $K$ of the continuous process,
which, for this simple example, can be computed analytically
from the expansion
\begin{equation*}
	 k(y,y') = \sum_{j=1}^\infty \tilde{\lambda}_j e_j(y)e_j(y'),
\end{equation*}
where $\tilde{\lambda}$ is an eigenvalue of $k(y,y')$ with eigenvector $e(y)$.
The eigenvalues and eigenvectors are defined by
\begin{equation*}
	\int_0^1 k(y,y')\,e(y)\;dy = \tilde{\lambda} \,e(y'),
\end{equation*}
and 
\begin{equation*}
	\int_0^1 e_i(y)e_j(y)\;dy =
	\left\{
	\begin{array}{l}
		1 \quad \text{ if } i = j,\\
		0 \quad \text{ otherwise.}
	\end{array}
	\right.
\end{equation*}
Thus,
\begin{align*}
	        \int_0^1 \int_0^1 k(y,y)^2 \; dy dy' &
		        = \sum_{j=1}^\infty \tilde{\lambda}_j^2
	= \pi\, L\left( \left(\exp\left(-\frac{1}{L^2}\right)-1\right) +\sqrt{\pi}\,\text{Erf}\left(\frac{1}{L}\right)\right)
\end{align*}
Since $\tilde{\lambda}\approx h\lambda$,
we have that
\begin{align*}
	 \vert\vert P\vert\vert_F^2 &\approx
	  \sum_{j=1}^\infty \tilde{\lambda}_j^2=  \pi\,\left( L\left( \exp\left(-\frac{1}{L^2}\right)-1\right) +\sqrt{\pi}\,\text{Erf}\left(\frac{1}{L}\right) \right).
\end{align*}
We find good agreement between the infinite dimensional results
and their finite dimensional approximations 
(the dashed lines are mostly invisible because they
coincide with the results calculated from the infinite dimensional problem).

The right panel of the figure shows the variation of $||P||_F$ and the dimension $\ell$
with the correlation length $L$.
We observe that 
the Frobenius norm remains almost constant for $L<10^{-2}$.
On the other hand, the dimension $\ell$ in~(\ref{effDim2})
increases as the correlation length decreases. What this figure shows is that the feasibility of
data assimilation depends on the level of noise,  but not on the effective number of
variables. One can find the same level of noise for very different effective dimensions. If data assimilation is not feasible, it is because the problem has too much
noise, not too many variables.

It is interesting to consider the limit $L \rightarrow 0$ in our family of processes. 
A stationary process $u(x)$ such that for every $x$, $u(x)$ is a Gaussian variable, with
$E[u(x)u(x')]=0$ for $x \ne x'$ and $E[u(x)^2]=A$, where $A$ is a finite constant, has very little
energy; white noise, the most widely used Gaussian process with independent values, has
$E[u(x)u(x')]=\delta_{x,x'}$, where the right-hand-side is a $\delta$ function, so that 
$E[u(x)^2]$ is unbounded. The energy of white noise is infinite, as is well-known.
If $k(x-x')$ in our family blew up like $L^{-1}$ at the origin as $L \rightarrow 0$,  the process
would be 
a multiple of Brownian motion; here it blows up like $L^{-1/2}$, allowing the energy to remain finite while not allowing the $E[u^2]$ to remain bounded. The moral of this discussion
is that a sampling problem where $P=I_m$ for all $m$ is unphysical.

An apparent paradox appears when one applies
our theory to determine the feasibility of a Gaussian
random variable with covariance matrix $P=I_m$, as in\cite{Bickel,BickelBootstrap,Bickel2,Snyder,SBM15}.
In this case,  $\vert\vert P\vert\vert_F =\sqrt{m}$,
so that the problem is infeasible by our definition if the number of variables $m$ is large.
On the other hand, the problem seems to be physically plausible.
For example, suppose one wishes to estimate the wind velocity at $m$ geographical sites
based on independent local models for each site
(e.g.~today's wind velocity is yesterday's wind velocity with some uncertainty),
and independent noisy measurements at each site.
The local $P_j$ at each site is one-dimensional and equals $1.$
Why then not try to determine the $m$ velocities in one fell swoop,
by considering the whole vector $x$ and setting $P=I_m$ ?
Intuition suggest that the set of local problems is equivalent to the $P=I_m$
problem, 
e.g.~ that one can use local stochastic models to predict the velocities at nearby sites,
and then mark these on a weather map and obtain a plausible global field.
Thus, $\vert\vert P\vert\vert_F$ is large in a plausible and feasible problem.
This, however, is not so. First, in reality, the uncertainties in the velocities at nearby sites are 
highly correlated, while the problem $P=I_m$ assumes that this is not so. The resulting velocities map from the latter will be unphysical and lead to wrong forecasts-  
 large scale coherent flows in today's weather map will have a different impact on tomorrow's forecast than a set of uncorrelated wind patterns, and of course carry a
much larger energy, even if the local amplitudes are the same. If one has a global problem
where the covariance matrix is truly $I_m$, then replacing the solution of the full problem
by a component-wise solution changes estimate of the noise from $||P||_F$ 
to 
the numerically smaller maximum of the component variances, which is not as
good a measure of the distance between the samples and their mean.  

\section{Linear analysis of the convergence of data assimilation}

We now summarize the linear analysis of the conditions under which particular particle filters produce
reliable estimates when the problem is linear (see\cite{CM13}).
A general nonlinear analysis is not within reach,
while the linear analysis captures the main issues.

The dynamical equation now takes the form:
\begin{equation}
\label{lindyn}
x^{n+1}=Ax^n+w^n,
\end{equation}
where $A$ is an $m\times m$ matrix
and, the data equation becomes
\begin{equation}
b^{n+1}=Hx^{n+1}+\eta^n,
\label{lindata}
\end{equation}
where $H$ is an $k \times m$ matrix.
As before, $w^n$ are independent $\mathcal{N}(0,Q)$
and the $\eta^n$ are independent $\mathcal{N}(0,R)$,
independent also of the $w^n$.
We assume that equation (\ref{lindyn}) is stable and
the data are sufficient for assimilation
(for a more technical discussion of these assumptions, see e.g. \cite{CM13}).
These two equations define a posterior probability density,
independently of any method of estimation.
The first question is, under what conditions is
state estimation with the posterior feasible in principle.
If one wants to estimate the state at time $n$,
given the data up to time $n$,
this requires that the covariance of $x^n\vert b^{1:n}$
have a small Frobenius norm.

The theoretical analysis of the Kalman filter\cite{Kalman1960} can be used to estimate this covariance,
even when the problem is not feasible in principle
or the Kalman filter itself is too expensive for practical use.
Under wide conditions,
the covariance of $x^n\vert b^{1:n}$
rapidly reaches a steady state
\begin{equation*}
	P=(I-KH)X,
\end{equation*}
where $X$ is the unique positive semi-definite solution of a particular nonlinear
algebraic Riccati equation (see e.g.\cite{Lancaster}) and
\begin{equation*}
	K=XH^T(HXH^T+R)^{-1}.
\end{equation*}
One can calculate $\vert\vert P\vert\vert_F$ and
decide whether the estimation problem is feasible in principle.
Thus, a condition for successful data assimilation is that
$\vert\vert P\vert\vert_F$ be moderate,
which generaly requires that the $||Q||_F,\,||R||_F$ 
in equations~(\ref{lindyn}) and~(\ref{lindata}) be small enough.      

A particle filter can be used to estimate the state by sampling
the posterior pdf $p(x^{0:n+1}|b^{1:n+1})$, defined by~(\ref{lindyn}) and~(\ref{lindata}).
The state at time $n$ conditioned on the data up to time $n$
can be computed by marginalizing samples of $p(x^{0:n+1}|b^{1:n+1})$
to samples of $p(x^n|b^{1:n+1})$ (which amounts to simply dropping the sample's past).
Thus, a particle filter does not directly sample
$p(x^n|b^{1:n+1})$, so that its samples carry weights (even in linear problems).
These weights must not vary too much or else the filter ``collapses'' (see section~2).
Thus, a particle filter can fail, even if the estimation problem is feasible in principle.

It was shown in\cite{CM13,Bickel,BickelBootstrap,Bickel2,Snyder}
that the variance of the negative logarithm of the weight must be small
or else the filter collapses.
For the SIR filter,
the condition that this collapse not happen is that the Frobenius norm of the 
matrix 
\begin{equation}
\Sigma_\text{SIR}=H(Q+APA^T)H^TR^{-1},
\label{SIR}
\end{equation}
be small enough.
For the optimal filter, which in the present linear setting coincides with the
implicit filter, success requires a small Frobenius norm for 
\begin{equation}
\label{OPT}
\Sigma_\text{Opt}=HAPA^TH^T(HQH^T+R)^{-1},
\end{equation}
see \cite{CM13}. In either case, this additional condition must be satisfied as well as the the condition
that $||P||_F$ is small. 

\begin{figure}[tb]
\centering\includegraphics[width=.6\textwidth]{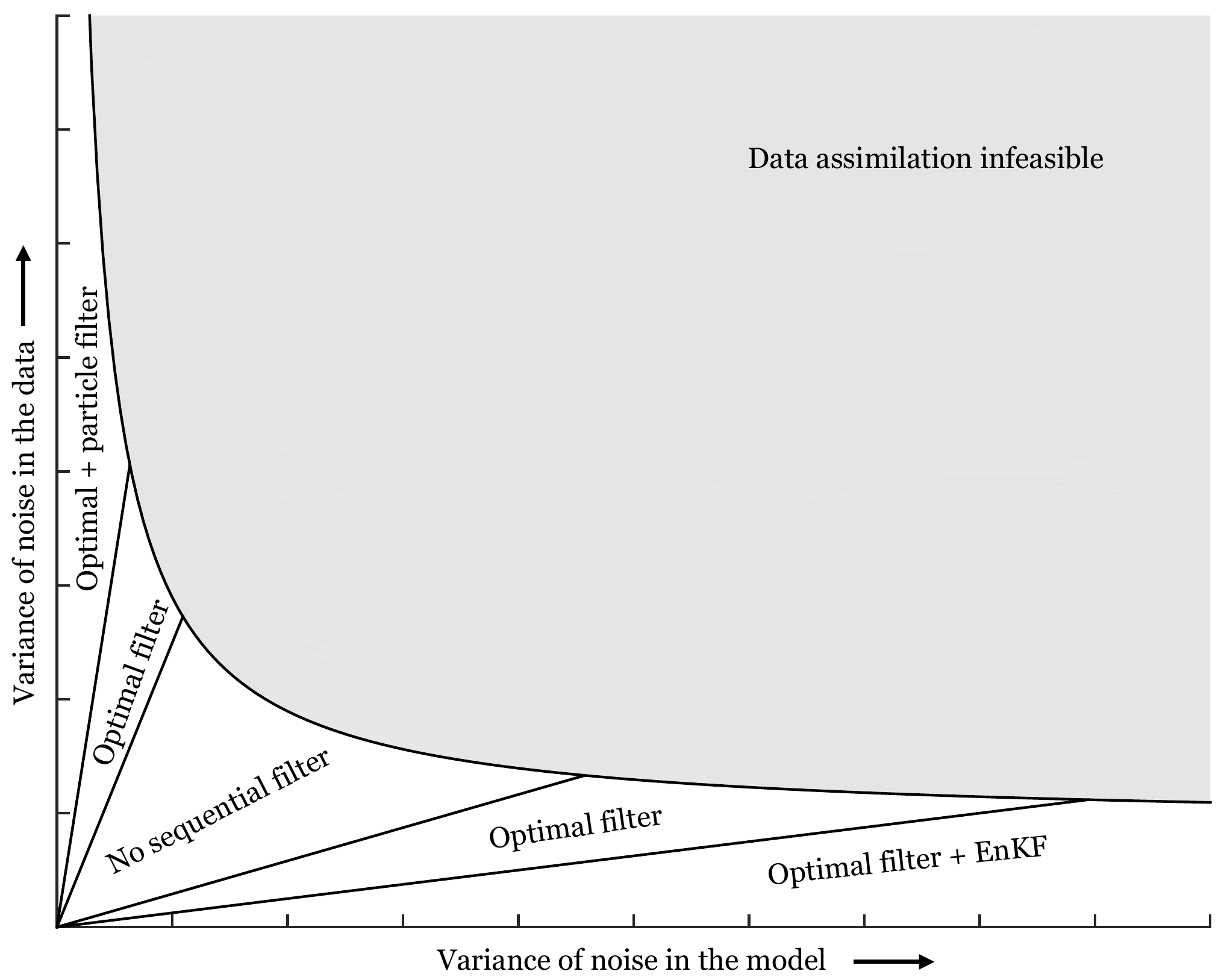}
\caption{Conditions for feasible data assimilation and
for successful sampling with the optimal particle filter, the particle filter and
the ensemble Kalman filter (from\cite{MH15}).}
\label{fig:Cones}
\end{figure}

To understand what these formulas say, we apply them to a simple model problem
which is popular as a test problem in the analysis of numerical weather prediction schemes. 
The problem is defined by
$H=A=I_m$ and $Q=qI_m,\,R=rI_m$,
where we vary the number of components $m$.
Note that for a fixed $m$, the problem is parameterized
by the variance $r$ of the noise in the model and the variance $q$ of the noise in the data.
This problem is feasible in principle if
\begin{equation*}
	\vert\vert P \vert\vert_F = \sqrt{m}\, \frac{\sqrt{q^2+4qr}-q}{2},
\end{equation*}
is not too large.
For a fixed $m$, this means that
\begin{equation*}
	\frac{\sqrt{q^2+4qr}-q}{2}\leq 1,
\end{equation*}
where the number $1$ stands in for a sharper estimate of
the acceptable variance for a given problem
(and this choice gives the complete qualitative story).
In figure~\ref{fig:Cones}, this condition is illustrated
and shown are all feasible problems in white and all
infeasible problems in grey.
The analysis thus shows that for data assimilation to be feasible,
either the model or the data must be accurate.
If both are noisy, the noise dominates so that estimation is useless.

In the SIR filter, the additional condition for success becomes
\begin{equation*}
\frac {\sqrt{q^2+4qr}+q}{2r} \le 1,
\end{equation*}
and for the optimal/implicit case, the additional condition is
\begin{equation}
\frac {\sqrt{q^2+4qr}-q}{2(q+r)} \le 1.
\end{equation}
In both cases, the number $1$ on the right-hand-side stands
in for a sharper estimate of the acceptable variance of the weights
for a given problem (which also depends on the computing resources).
In either case, the added condition is 
quadratic and homeogeneous in the ratio $q/r$,
and thus slices out conical regions from the region where data assimilation is feasible in principle.
These conical regions are shown in figure~\ref{fig:Cones}.
The region where estimation with a particular filter is feasible is labeled
with the name of the particular filter.
Note that we also show results for the ensemble Kalman filter (EnKF)\cite{EvensenBook,Tippet2003},
which are derived in \cite{MH15}, but not discussed in the present paper.
The analysis of the EnKF relates to the situation where it is used
to sample the same posterior density as the other filters quoted in the figure.
The analysis in~\cite{MH15} also  includes the situation where the EnKF is used to
sample a marginal of that pdf directly, when the conclusions are different.

In practical problems one usually tries to sharpen estimates obtained from
approximate dynamics with the help of accurate measurements,
and the optimal\slash implicit filter works well in such problems.
However, there is a region in which not even the optimal filter succeeds.
What fails there is 
the sequential approach to the estimation problem,
i.e.~the factorization of the importance function as in~(\ref{factor}),
see\cite{MH15,SBM15}.
Non-recursive filters can succeed in this region,
and there too implicit sampling can be helpful
i.e.~one can apply implicit sampling direction to $p(x^{0:n}\vert b^{1:n})$.

One conclusion we reach from our analysis is that if Bayesian estimation
is feasible in principle then it is feasible in practice. However, it is assumed in the previous analysis so far that one
has suitable priors, or equivalently, one knows what the noise $w^n$ in equation (\ref{sde}) really is.
The fact is that generally one does not. 
We discuss this issue in the next section.

\section{Estimating the prior}

The previous sections described how one may use a prior and a distribution of observation errors
to estimate the states or the parameters of a system. This estimation depends on knowing the prior. As we have seen, in data assimilation, the prior is
generated by the solution of an equation such as 
the recursion (\ref{sde}) and depends on knowing the noise $w^n$ in that equation. It is often assumed that the noise in that equation is white. However, one can show that the noise is not white in most problems (see e.g. \cite{Wil05,CVE08,CH14}). We now present a preliminary 
discussion of methods for estimating the noise. 

First, one has to make some assumptions about the origin of the noise. A
reasonable assumption (see e.g. \cite{CH14, BH14}) is that there is noise in the equations of motion because
a practical
description of nature is necessarily incomplete. For example, one can write a solvable
equation of motion for a satellite turning around the earth by assuming that
the gravitational pull is due to a perfectly spherical earth with a density
that is a function only of distance from the center (see \cite{MG15}). Reality is different,
and the difference produces noise, also known as model error. The problems
to be solved are: (i) estimate this difference, (ii) identify it, i.e.,
find a concise approximate representation of this difference that can be
effectively evaluated or sampled on a computer, and (iii) design an
algorithm that imbeds the identified noise in a practical
data assimilation scheme. These problems have been discussed in a number of recent papers, e.g. \cite{CH14,MH13},
in particular the review paper \cite{Har13} which contains an extensive
bibliography.

Assume that the full description of a physical system has the form:
\begin{equation}
\frac {d}{dt}x=R(x,y),\,\,\,\, \frac{d}{dt}y=S(x,y),
\label{main}
\end{equation}
where $t$ is the time, $x=(x_1,x_2,\dots,x_m)$ is the vector of resolved variables, and $y=(y_1,y_2,\dots,y_{\ell})$ is the vector of unresolved variables, with initial data $x(0),y(0).$ Consider a situation where this system is too complicated to solve, but where data are available, usually as
sequences of measured values of $x$, assumed here to be observed with negligible observation errors.
Write $R(x,y)$ in the form
\begin{equation}
R(x,y)=R_0(x)+z(x,y),
\label{reduce}
\end{equation}
where $R_0$ is a practical approximation of $R(x,y)$ and one is able to solve the equation
\begin{equation}
\frac {d}{dt}x=R_0(x).
\label{Rzero}
\end{equation}
However, $x$ does not satisfy equation (\ref{Rzero}) because the true equation, the first of equations (\ref{main}), is more complicated. The difference between the true equation and the approximate equation is the remainder 
$z(x,y)=R(x,y)-R_0(x)$, which is the noise in the determination of $x$. In general, $z$ has to be determined from
the data, i.e., from the observed values of $x$. 

A usual approach to the problem of estimating $z$ is to use equation (\ref{reduce}) to obtain its values from $x$ data, i.e. calculate $z=\frac {d}{dt}x-R_0(x)$, and then identify it as a continuous stochastic process.
This may be difficult, in particular, calculating $z$ requires that one differentiate $x$,
which is generally impractical or inaccurate because $z$ may
have high-frequency components or fail to be sufficiently smooth, and the data may not be available at sufficiently small time intervals (an illuminating analysis in a special case can be found in \cite{PSW09,ST12}).  Once one has
values of $z$, identifying it as a function of the continuous variable $t$ 
often requires making unwarranted assumptions on the small-scale structure of
$z$, which may be unknown when the data are available only at discrete times.

An alternative is supplied by a discrete approach as in \cite{CL15}. Equation (\ref{Rzero}) is always solved on the computer, i.e., in discrete form, the data are always given at discrete
points, and it is $x$ one wishes to determine
but in general one is not interested in determining $z$ per se. We can therefore avoid the difficult detour
through a continuous-time $z$ followed by a discretization, as follows.
We pick once and for all a particular discretization of equation
(\ref{Rzero}) with a particular time step time
step $\delta$,
\begin{equation}
x^{n+1}=x^n+\delta R_{\delta}(x^n),
\label{discrete}
\end{equation}
where $R_{\delta}$ is obtained, for example, from 
a Runge--Kutta method, and where $n$ indexes the result after $n$ steps;
the differential equation has been reduced to a recursion such as equation (\ref{sde}).
We then use the data to identify the discrepancy sequence, $z_{\delta}^{n+1}=(x^{n+1}-x^n)/\delta-R_{\delta}(x^n)$, which is available from $x$ data without approximation.

Then assume, as one does in the continuous-time case, that the system under consideration is ergodic, so that its long-time statistics are stationary. The sequence $z_{\delta}^n$ becomes a stationary time series,
which can be represented by one of the representations of time series, e.g.
the NARMAX (nonlinear auto-regression moving average with exogenous inputs) representation, with $x$ as an exogenous input. The observed $x$ of course may include observation errors, which have to be separated from the model noise via
some version of filtering.

\begin{figure}[tb]
\centering\includegraphics[width=.5\textwidth]{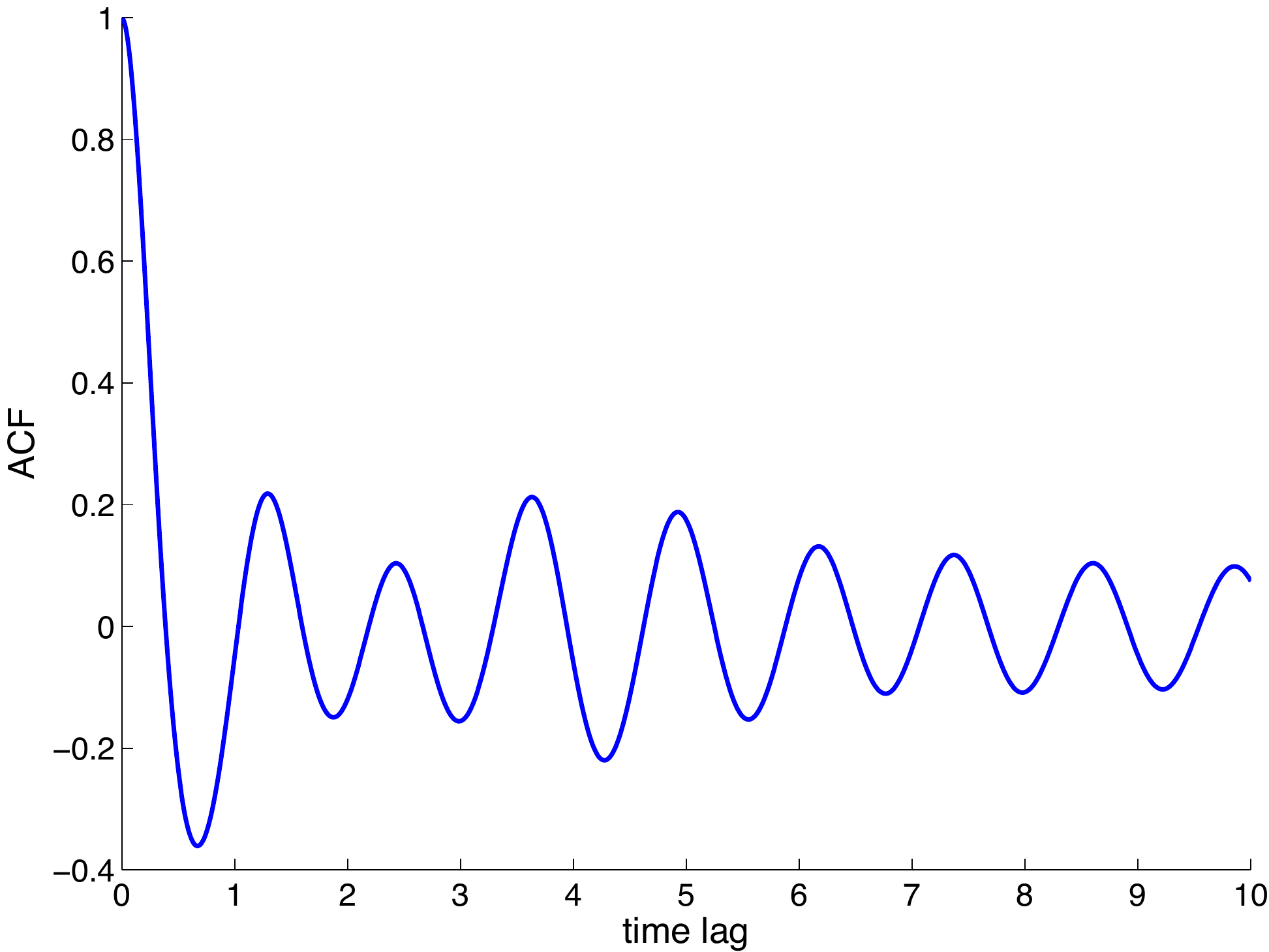}
\caption{Autocorrelation function of the noise $z$ in the Lorenz 96 system.}
\label{fig:zACF}
\end{figure}

As an example, we applied this construction to the Lorenz 96 system \cite{Lor95}, created to serve as a metaphor for the atmosphere, which has
been widely used as a test bench for various reduction and filtering schemes.
It consists of a set of chaotic differential equations, which, following
\cite{FVE04},
we write as:
\begin{eqnarray}
\frac{d}{dt} x_k & =x_{k-1}\left( x_{k+1}-x_{k-2}\right) -x_{k}+F+z_{k},
\label{L96_x} \\
\frac{d}{dt} y_{j,k}& =\frac{1}{\varepsilon }%
[y_{j+1,k}(y_{j-1,k}-y_{j+2,k})-y_{j,k}+h_{y}x_{k}] \notag
\end{eqnarray}
with
\begin{equation*}
z_{k}=\frac{h_{x}}{J}\sum_{j}y_{j,k},
\end{equation*}%
and $k=1,\dots ,K$, $j=1,\dots ,J$.
The
indices are cyclic,
$x_{k}=x_{k+K},~y_{j,k}=y_{j,k+K}$ and $y_{j+J,k}=y_{j,k+1}$. This 
system is invariant under spatial translations, and the statistical
properties are identical for all $x_{k}$.
The parameter $\varepsilon$ measures the time-scale separation
between the resolved variables $x_{k}$ and the unresolved variables $y_{j,k}$.
The parameters are $\varepsilon=0.5,$ $K=18,J=20,F=10,h_{x}=-1$ and $h_{y}=1$. The ergodicity of the Lorenz 96 system has been established numerically in earlier work (see e.g. \cite{FVE04}). We pretend that this system is too difficult to integrate
in full; we take as as $R_0$ of equation (\ref{Rzero}) the system is which  is $z_k=0$. The noise is then $z_k$, which in our special case can actually be calculated by solving the full system of equations; in general the noise has to be estimated from data as described above and in \cite{Wil05, CL15}. 
In figure~\ref{fig:zACF} we plot the covariance function of the noise component $z_1$ determined as just described. Note that $z$ cannot be thought of white noise even remotely. To the extent that the Lorenz 96 is a valid metaphor for 
the atmosphere, we find that the noise in a realistic problem is not white. 
This can of course be deduced from the construction after equations (\ref{main}), where the noise appears as the difference between solutions of differential equations, which is not likely to be white. 

This relation between the preceding discussion of the noise (which defines the prior through equation (\ref{reduce})) should
be compared with the discussion of implicit sampling and particle filters. Here too the data have been put to additional
use (there to define samples and not only to weight them, here to provide information about the noise as well as
about the signal); here too an assumption of a white noise input has been found wanting, and realism requires
non-trivial correlations, (there in space, here in time).

The next step is to combine the recursion (\ref{discrete})
with observations to produce a state estimate. Recent work
on this topic is summarized in \cite{Har13}. An example of this construction  
could not be produced in time to appear in this article. 

\section{Conclusions}
We have presented algorithms for data assimilation and for estimating the prior, 
with some analysis. The work presented shows that the keys to success are a better
use of the data and a careful analysis of what is physically plausible and useful.
We feel that significant progress has been made. One conclusion we draw from this 
work is that data assimilation, which is often considered as a problem in statistics,
should also, or even mainly, be viewed as a problem in computational physics.
This conclusion has also been reached by others, see e.g. \cite{MH13,CVE08}.

\section{Acknowledgements}
This work was supported in part by the Director, Office of Science, Computational and Technology Research, U.S. Department of Energy, 
under Contract No. DE-AC02-05CH11231, and by the National Science
Foundation under grants DMS-1217065 and DMS-1419044.

\bibliographystyle{plain}
\bibliography{References}

\end{document}